\documentclass[12pt,graphics]{iopart}

\newcommand{\Journal}[4]{{#1} #4 {\bf #2}, #3 }
\usepackage{iopams}  
\usepackage{graphicx}
\usepackage{marvosym}        
\usepackage{epsfig}
\usepackage{units}
\usepackage{multicol,graphics,verbatim}
\usepackage{longtable}
\usepackage{colortbl}
\usepackage{color}
\usepackage{rotating}  
\usepackage{nicefrac} 
\usepackage{calc} 
\usepackage{amssymb}
\usepackage{amsthm}
\usepackage{upgreek}          
\usepackage{ifthen}
\usepackage[colorinlistoftodos]{todonotes}
\usepackage{lineno}

\usepackage{threeparttable}

\newcommand{\NIMA}{{\em Nucl. Instrum. Methods} A}

\newcommand{\NPB}{{\em Nucl. Phys.} {\bf B}}

\newcommand{\NPA}{{\em Nucl. Phys.} A}
\newcommand{\PLB}{{\em Phys. Lett.}  {\bf B}}
\newcommand{\PRD}{{\em Phys. Rev.} {\bf D}}
\newcommand{\PRC}{{\em Phys. Rev.} C}

\newcommand{\ZPA}{{\em Z. Phys.} A}

\newcommand{\PAN}{\em Phys. Atom. Nucl.}

\newcommand{\IJARI}{\em Int. J. Appl. Radiat. Isot.}
\newcommand{\NAT}{\em Nature}

\newcommand{\bpbp}{\mbox{$\beta^+\beta^+$} }
\newcommand{\ecec}{\mbox{$ECEC$} }
\newcommand{\bec}{\mbox{$\beta^+EC$} }
\newcommand{\bnel}{\mbox{$\bar{\nu}_e$} }

\newcommand{\obb}{0\mbox{$\nu\beta\beta$ decay}} 
\newcommand{\zbb}{2\mbox{$\nu\beta\beta$ decay}}

\newcommand{\nel}{\mbox{$\nu_e$}}

\newcommand{\gess}{\mbox{$^{76}$Ge }}

\newcommand{\gevs}{\mbox{$^{74}$Ge}}
\newcommand{\sevs}{\mbox{$^{74}$Se}}

\newcommand{\xehs}{\mbox{$^{136}$Xe }}

\newcommand{\tzn}{\mbox{T$_{1/2}^{2\nu}$} }

\newcommand{\grays}{$\gamma$-rays}
\newcommand{\gray}{$\gamma$-ray}
\newcommand{\glines}{$\gamma$-lines}
\newcommand{\gline}{$\gamma$-line}
\newcommand{\be}{\begin{equation}}
\newcommand{\ee}{\end{equation}}
\def\bea{\begin{eqnarray}} 
\def\eea{\end{eqnarray}} 
\newcommand{\ra}{\rightarrow }

\begin{document}

\today

%
%
%
\title{Double electron capture searches in $^{74}$Se}

\author{B. Lehnert$^a$, T. Wester$^a$, D. Degering$^b$, D. Sommer$^a$, L. Wagner$^{a,c}$, K. Zuber$^a$}
\address{$^a$ Institut f\"ur Kern- und Teilchenphysik, Technische Universit\"at Dresden,\\
Zellescher Weg 19, 01069 Dresden, Germany\\
$^b$ VKTA - Strahlenschutz, Analytik \& Entsorgung Rossendorf e.V., 01314 Dresden, Germany\\
$^c$ Helmholtz-Zentrum Dresden-Rossendorf (HZDR), Bautzner Landstr. 400, 01328 Dresden, Germany}
\ead{bjoernlehnert@gmail.com, thomas.wester@tu-dresden.de,\\ 
detlev.degering@vkta.de, zuber@physik.tu-dresden.de}

\begin{abstract}
A search for various double electron capture modes of $^{74}$Se has been performed using
an ultralow background Ge-detector in the Felsenkeller laboratory, Germany. Especially for the potentially resonant transition into the \unit[1204.2]{keV} excited state of $^{74}$Ge a lower half-life limit of \unit[$0.70\cdot 10^{19}$]{yr} (90 \% credibility) has been obtained. 
Serious concerns are raised about the validity of obtained $^{74}$Se limits in some recent publications.

\end{abstract}

\maketitle

\section{Introduction}
The search for physics beyond the standard model is a wide spread
activity in accelerator and non-accelerator physics. Among all the searches
for new physics, total lepton number violation plays 
an important role. 
The golden channel to search for total lepton number violation is 
neutrinoless double beta decay 
\be
(Z,A) \ra (Z+2,A) + 2 e^-  \quad (\obb).
\ee   

For recent reviews see \cite{avi08,Rodejohann:2011fr,Rodejohann:2012cc}. Any Beyond Standard Model (BSM) physics allowing $\Delta L =2$ processes can contribute to the decay rate. It has been shown \cite{Schechter:1982} that its observation would imply that 
neutrinos are their own antiparticles (Majorana neutrinos) which is an
essential ingredient for leptogenesis, explaining the
baryon asymmetry in the universe with the help of Majorana neutrinos 
(see for example \cite{fon12}). However, it is unknown how much individual BSM processes contribute to the double beta decay rate. 

To observe $0\nu\beta\beta$ decay, single beta decay has to be forbidden by energy conservation or at least strongly suppressed. For this reason only 35 potential double beta minus emitters exist. As the phase space for these decays
scales strongly with the Q-value, searches are using only those nuclides with a Q-value above
\unit[2]{MeV}, reducing the list to 11 candidates. Lower limits on half-lives beyond $10^{25}$ years of the neutrino less mode have been measured for the isotopes
\gess and \xehs  \cite{gan13,ago13,alb14}. 

In addition, the allowed  process of neutrino accompanied double beta decay 
\be
(Z,A) \ra (Z+2,A) + 2 e^-  + 2 \bnel \quad (\zbb)
\ee      
will occur. It is the rarest decay measured in nature and has been observed 
in more than ten isotopes.
However, in both cases the measured quantity is a half-life, which is linked to the phase space $G$ and nuclear transition
matrix elements $M$. In case of \zbb\ the matrix element is purely Gamow-Teller (GT) and the relation is 
\be
\left(\tzn\right)^{-1} = G \times \mid M_{GT} \mid^2\,,
\ee
which does not require any BSM particle physics, as opposed to \obb. 
The half-life measurements 
of \zbb\ is important for understanding the nuclear structure since it will provide valuable information on the nuclear matrix elements which can be directly compared with theory.

Equivalent processes to double beta minus decay with the emission of two electrons could occur on the right side of the
mass parabola of even-even isobars. 34 nuclides are candidates for this process. Three different decay modes are possible involving $\beta^+$ decay and
electron capture (EC)
\begin{eqnarray}
\label{eq:ecec}
(Z,A) &\ra  (Z-2,A) + 2 e^+ \; (+ 2 \nel) \quad & \mbox{(\bpbp)} \label{eqn:b+b+}\\
e^- + (Z,A) &\ra (Z-2,A) + e^+ \; (+ 2 \nel) \quad & \mbox{(\bec)} \label{eqn:ecec}\\
2 e^- + (Z,A) &\ra (Z-2,A) \; (+ 2 \nel) \quad & \mbox{(\ecec)} \label{eqn:ecb+}
\end{eqnarray}
Decay modes containing an EC emit X-rays or Auger electrons created by the atomic shell vacancy in the daughter nuclide. 
Decay modes containing a positron have a reduced Q-value as each generated positron accounts for a reduction 
of 2 $m_ec^2$ in the phase space.
Thus, the largest phase space is available in the \ecec mode and makes it the most probable one. However, the \ecec is also the most difficult to detect, only producing X-rays (or Auger electrons) and two neutrinos in the final state instead of \unit[511]{keV} \grays~
resulting from the decay modes involving positrons. Furthermore, it has been shown that \bec transitions have an enhanced sensitivity to right-handed weak currents (V+A interactions) \cite{hir94} and thus would help to disentangle the physics mechanism of \obb, if observed.

In the 0$\nu$\ecec mode there are only X-rays (or Auger electrons) and if there is no other particle in the final state this would violate energy and momentum conservation. It was suggested that the energy is released radiatively as a single internal radiative bremsstrahlung \gray, two \grays, an $e^-e^+$-pair or an internal conversion electron \cite{Vergados:1983do,Doi:1993hj}.
In this case the rate is reduced by several orders of magnitude due to the additional radiative coupling and half-lives in the order of $10^{29}$ to \unit[$10^{32}$]{yr} have been predicted for various isotopes assuming an effective neutrino mass of \unit[1]{eV} \cite{Sujkowski:2004dub}.
However, a potential resonance enhancement for 0$\nu$\ecec is expected for isotopes in which the final state of the daughter nuclide is energetically degenerated with the ground state of the mother within a few \unit[100]{eV}. This may lead to an up to $10^6$ times faster rate \cite{Sujkowski:2004dub,ber83,Kotila:2014ira}.

\begin{figure}
\centering
\includegraphics[width=0.9\columnwidth]{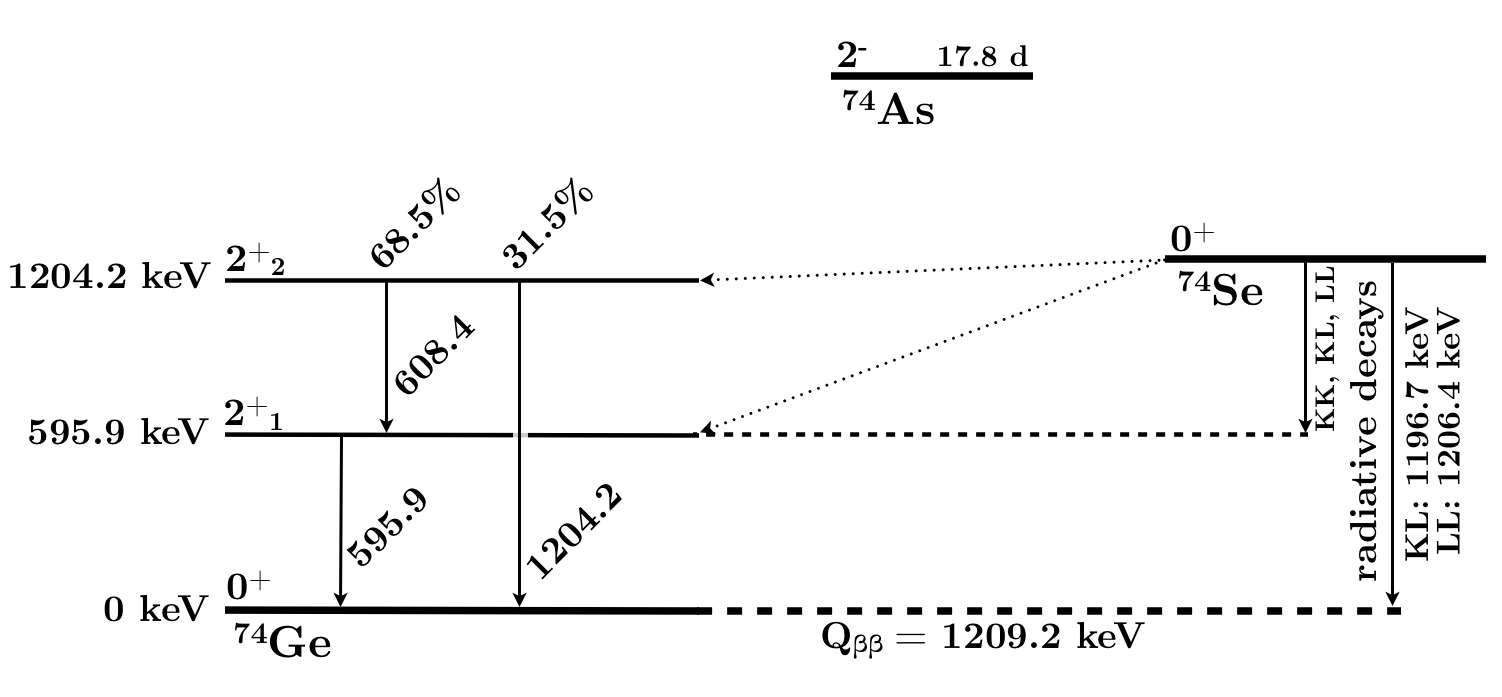}
\caption{\ecec decay scheme of \sevs. The radiative decays are separated into captures from the K and L atomic shells and their combination KK, KL and LL.}
\label{fig:decayscheme}
\end{figure}

This paper includes searches for the (radiative) 0$\nu$\ecec of \sevs\ into the ground state of \gevs\ and into the $2_1^+$ excited state, as well as for the 2$\nu$\ecec decay into the two excited states ($2_1^+$ and $2_2^+$).
An illustration of the different decay modes is shown in Fig.~\ref{fig:decayscheme}.
Depending on the shells the electrons are captured from, the energy of the \gray\ emitted in the radiative 0$\nu$\ecec decay modes can vary by a few keV. The dominant orbital is always the s-orbital in a given shell due to its finite probability of presence at the nucleus.
This work investigates several decay modes with captures from the s-orbital of the K and L shells. The energy of the X-rays (or Auger electrons) is 11.10~keV due to a vacancy in the K shell and 1.41~keV due to a vacancy in the L shell \cite{Nist:2014}. The Coulomb interaction between the two holes slightly changes the energy of the system compared to two individual single electron capture cases. This has been quantitative estimated in Ref.~\cite{Krivoruchenko2011} for a variety of double electron capture candidates and is typically well below \unit[1]{keV}. This effect is well included in the systematic uncertainty of the energy calibration of the Ge-detector.

Additionally, the $2^+_2$\ state at \unit[1204.205(7)]{keV} has been considered for a possible resonance enhancement. This decay mode was investigated in the past in Ref.~\cite{bara07,fre11} and most recently in Ref.~\cite{fre15}.
Unfortunately, a more precise Q-value measurement of \unit[1209.240(7)]{keV} \cite{kol10,mou10} seems to disfavor a non-radiative resonant transition and a new half-life of \unit[$5\cdot 10^{43}$]{yr} with \unit[1]{eV} effective neutrino mass has been calculated \cite{kol10}.
Recent measurements have set limits of $0.55\cdot 10^{19}$ yr \cite{bara07} and $1.5\cdot 10^{19}$ yr \cite{fre11,fre15}, though the latter limit seems unrealistic when comparing it to the sensitivity of the experiment, as shown later.

The de-excitation of the $2_2^+$ state can follow two branches (Fig.~\ref{fig:decayscheme}).
In the first branch with (68.5$\pm$1.4)\% probability two \grays\ with energies of \unit[595.9]{keV} and \unit[608.4]{keV} are emitted.
In the second branch 
only one \gray\ with an energy of \unit[1204.3]{keV} is emitted \cite{NuclData06}.
Hence, the potential signal from the non-radiative resonant 0$\nu$\ecec transition is expected to produce three peaks in the energy spectrum that correspond to the three \grays.
The third peak corresponding to \unit[1204.2]{keV} \gray\ additionally includes a small contribution of less than 10\% from the summation of the \unit[595.9]{keV} and \unit[608.4]{keV} \grays.

\section{Experimental Setup and Data}
\label{setup}

\begin{figure}
\centering
\includegraphics[width=0.7\columnwidth]{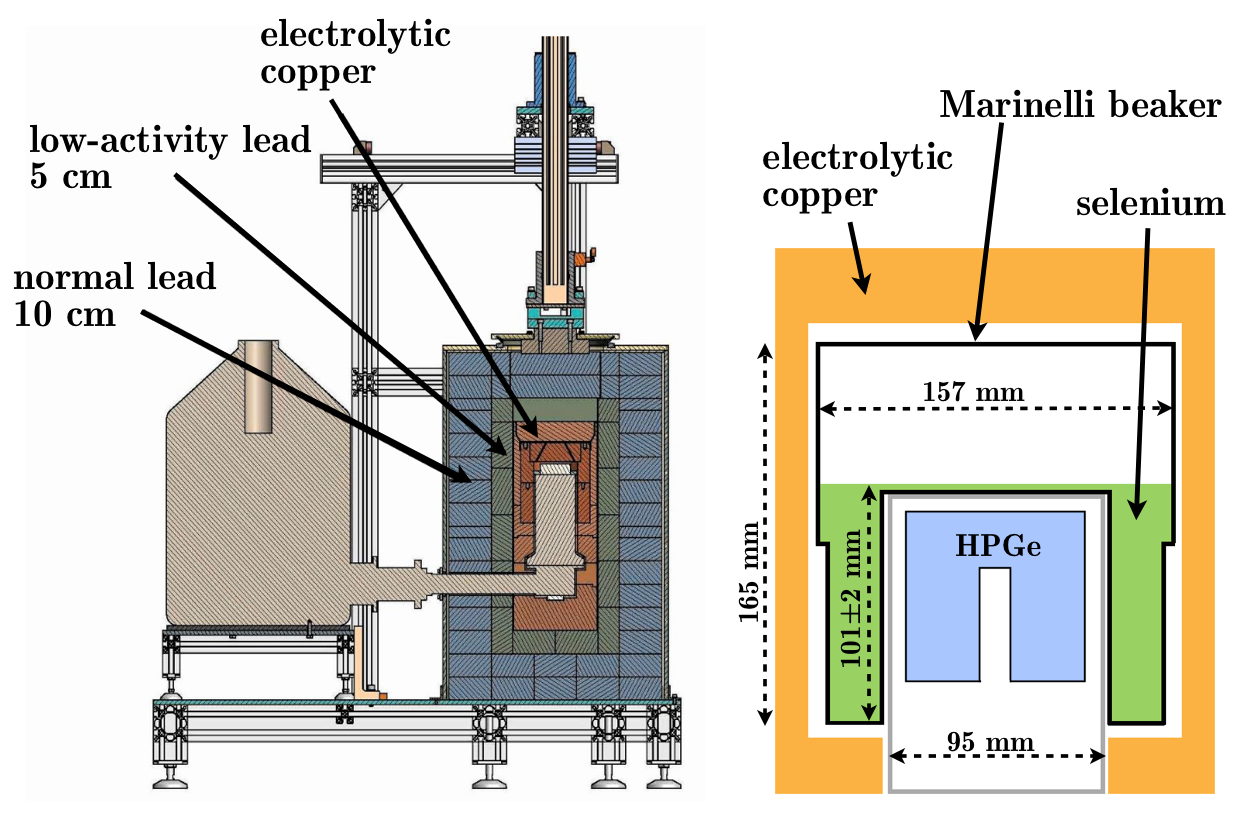}
\caption{Drawing of the setup, consisting of a HPGe detector surrounded by a copper and lead shield. The selenium sample was filled into a Marinelli beaker which fits onto the endcap of the detector.}
\label{fig:DetectorAndSetup}
\end{figure}

The measurement was performed in the Felsenkeller Underground Laboratory in Dresden, Germany, with a shielding of \unit[110]{m.w.e.}\ rock overburden reducing the muon flux to \unit[${0.6}\cdot 10^{-3}$]{cm$^{-2}$s$^{-1}$} \cite{Niese98}. 
A sample of \unit[2503.6]{g} selenium grains was used which is a large subset of the sample used in the measurement of Ref.~\cite{fre11,fre15}. \sevs\ has a natural abundance of 0.89\% translating into an isotopic mass of \unit[20.9]{g} of \sevs\ within the sample.
The sample was filled into a standard Marinelli beaker with an inner recess fitting onto the end cap of an ultra low background HPGe detector with a relative efficiency of \unit[90]{\%} routinely used for gamma spectroscopy measurements.
A schematic drawing of the arrangement can be seen in Fig.~\ref{fig:DetectorAndSetup}.
The detector is surrounded by a \unit[5]{cm} copper shielding embedded in another shielding of \unit[15]{cm} of low activity lead.
The inner \unit[5]{cm} of the lead shielding has a specific activity of \unit[($2.7 \pm 0.6$)]{Bq/kg} $^{210}$Pb while the outer \unit[10]{cm} has \unit[($33 \pm 0.4$)]{Bq/kg}.
The spectrometer is located in a measuring chamber which is an additional shielding against radiation from the ambient rock.
Furthermore, the detector is constantly held in a nitrogen atmosphere to avoid radon.
The data is collected with a 16384 channel MCA from ORTEC recording energies up to \unit[2.8]{MeV} resulting in a calibration slope of about \unit[0.17]{keV/channel}. 
The energy resolution is 0.24\% FWHM@\unit[595]{keV} and 0.15\% FWHM@\unit[1204]{keV}.
More details can be found in \cite{deg09,degering08}.
The sample was measured for 35.29 days corresponding to an exposure of \unit[88.35]{kg$\cdot$d}.

The efficiency calibration was performed by a mixture of analytically pure SiO$_2$, the reference materials RGU and RGTh from IAEA \cite{RGU} and KCl.
Those activity standards contain specific activities of \unit[($107.0\pm0.3$)]{Bq/kg} $^{238}$U, \unit[($113.0\pm1.2$)]{Bq/kg} $^{232}$Th and \unit[($106.9\pm2.1$)]{Bq/kg} $^{40}$K respectively.
The calibration sources were filled in an identical container as used for the selenium sample.
Thus, calibration source and measuring sample differ only in their self absorption behaviour.

The full energy detection efficiencies for \grays\ from the selenium sample was determined with Monte Carlo (MC) simulations based on Geant4.
The detector geometry was implemented in the code framework MaGe \cite{Boswell:hc} developed for particle propagation at low energies.
Events of the signal process were generated using a modified version of Decay0 \cite{decay0}
with updated branching ratios for the $2^+_2$ state and including the angular correlation between the emitted \grays.
The code was validated with the calibration source in the same geometry as the selenium sample.
For \glines\ in the range of \unit[$0.5-1$]{MeV}, the relative difference of detection efficiency between MC and calibration measurement is on average less than 10\%.
For the resonant decay to the $2^+_2$ state the efficiencies for \grays\ with energies of \unit[595.9]{keV} and \unit[1204.2]{keV}
were determined as \unit[($1.85 \pm 0.19$)]{\%} and \unit[($0.75 \pm 0.08$)]{\%} relative to one decay of a source nuclide including the branching ratio and summation effects. 
Contributions to the uncertainty come from the MC statistics (1\%), systematic uncertainties of MC processes and geometries (10\%) and the packing density of the selenium sample (3\%).

\begin{figure}
\centering
\begin{minipage}[b]{1\textwidth}
\centerline{\includegraphics[width=1\textwidth]{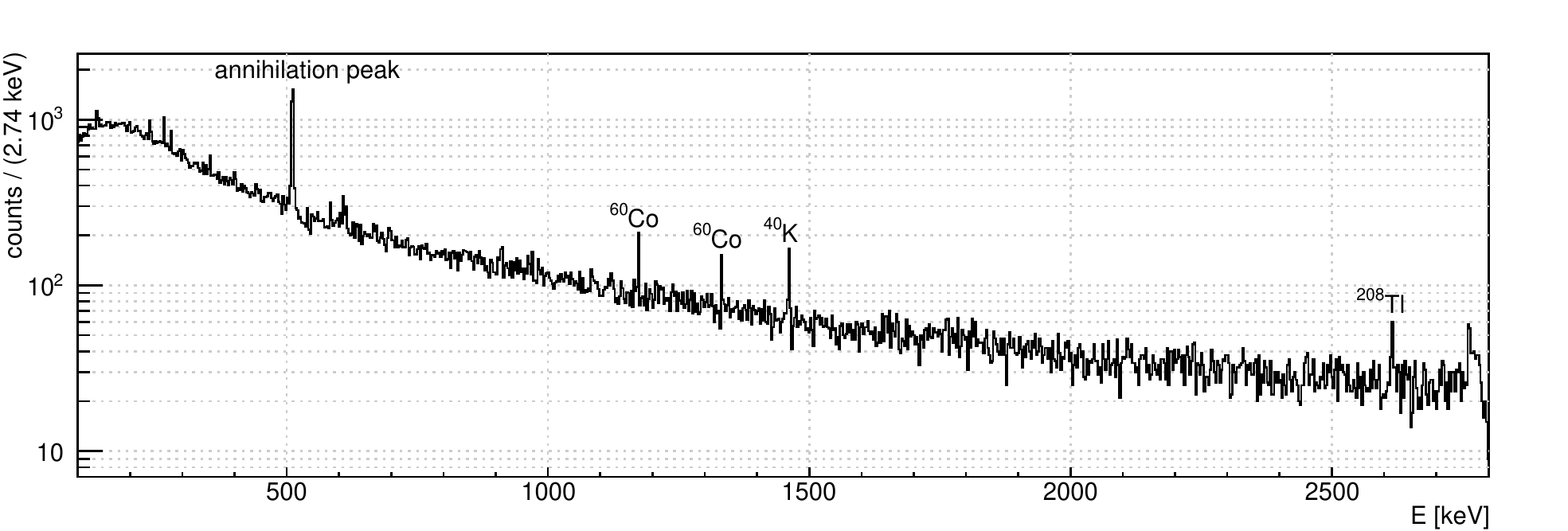}}
\end{minipage}%
\vspace{1mm}
\begin{minipage}[b]{1\textwidth}
\includegraphics[width=0.50\textwidth]{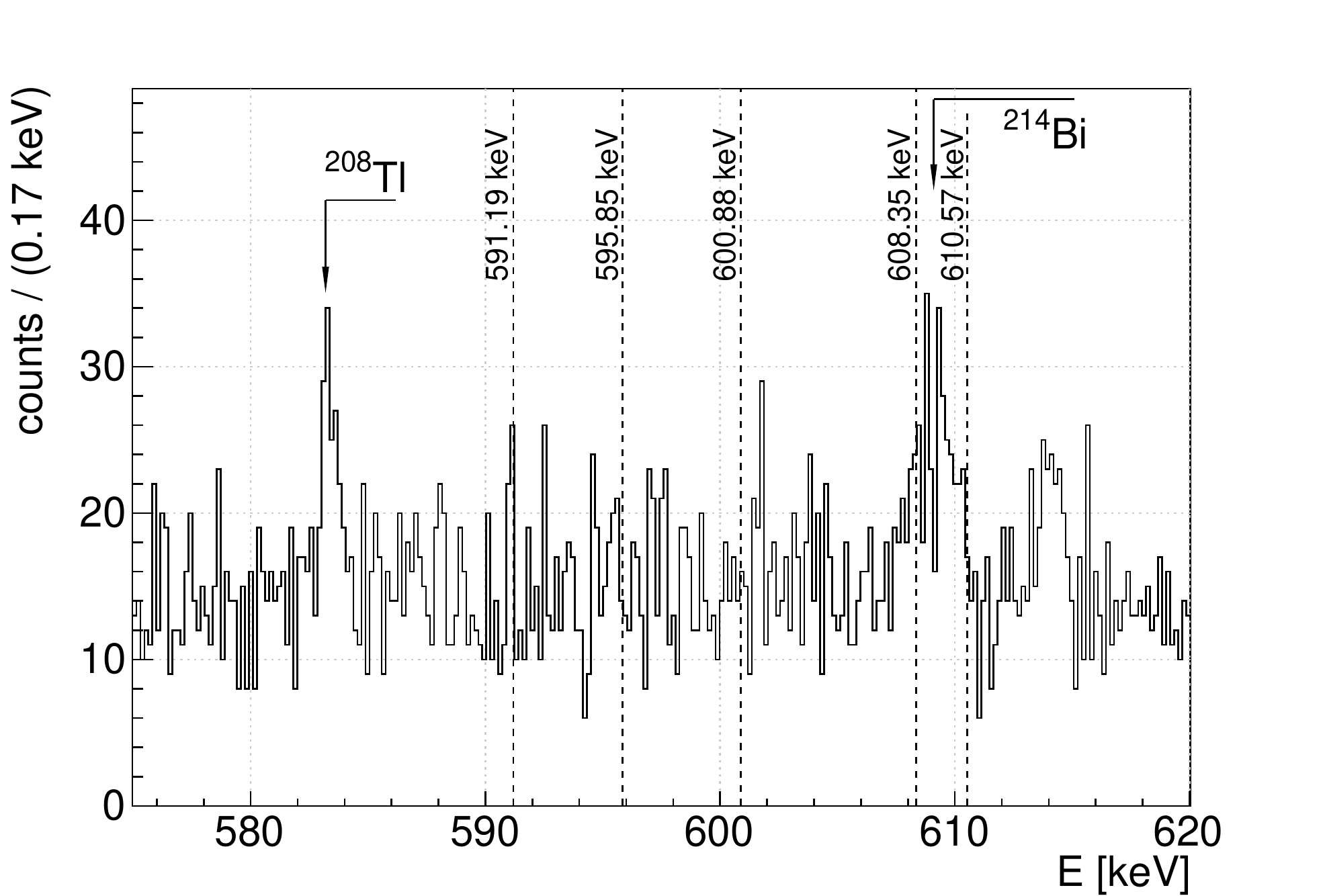}%
\includegraphics[width=0.50\textwidth]{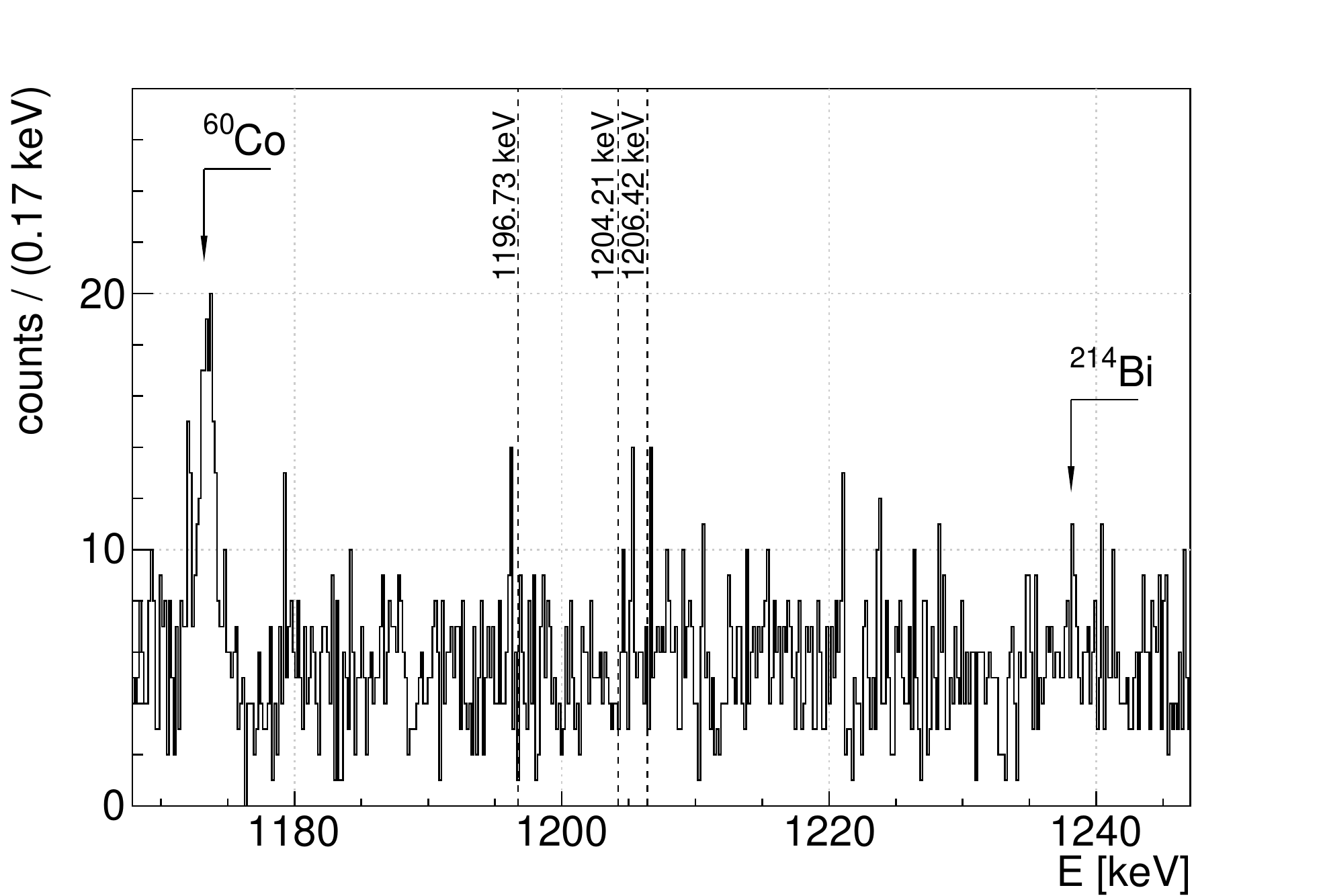}%
\end{minipage}%
\caption{Energy spectrum after 35.29 days measuring time.
The upper figure shows the spectrum in the full range up to 2.8~MeV. Background \glines\ from e$^+$e$^-$-annihilation, $^{60}$Co, $^{40}$K and $^{208}$Tl can be identified. In the bottom figures the spectrum is zoomed into the two regions of interest. Background \glines\ expected from $^{214}$Bi, $^{208}$Tl and $^{60}$Co are marked with arrows. The dashed lines indicate the position of \glines\ expected from the various \ecec decay modes of \sevs.}
\label{fig:EnergySpectrum}
\end{figure}

The measured energy spectrum can be seen in Fig.~\ref{fig:EnergySpectrum}.
The background \glines\ in the regions of interest (ROI) around \unit[595.9]{keV} and \unit[1204.2]{keV} are at \unit[609.3]{keV} and \unit[1238.1]{keV} from $^{214}$Bi and at \unit[583.2]{keV} $^{208}$Tl.
Especially the \unit[609.3]{keV} background \gline\ is very close to the \unit[608.4]{keV} signal \gline\ and  can not be separated with the given energy resolution. Hence, this \gline\ was not considered for analysis.

\section{Analysis}
\label{sec:Ana}
A Bayesian analysis with the help of the Bayesian Analysis Toolkit (BAT) \cite{BAT09} is used to extract information about the signal count expectation for the various decay modes from the spectrum. The analysis is described exemplary for the resonant 0$\nu$\ecec into the 2$^+_2$ state. Results for the other decay modes are obtained analogously.

\subsection{Fit procedure}
\label{sec:proflike}
A combined spectral fit of the energy spectrum in the two regions of interest (ROI) around the
\unit[595.9]{keV} and \unit[1204.3]{keV} \glines\ is performed.
The data is binned with a bin width of about \unit[0.17]{keV}, which corresponds to the width of the MCA channels.
Both fit regions are chosen so that no background peaks are included. 
The first ROI is defined as [585, 607] keV, limited below by the \unit[583.2]{keV} peak from $^{208}$Tl and above by the \unit[609.3]{keV} peak from $^{214}$Bi.
The second ROI is defined as [1176, 1235] keV, limited below by the \unit[1173.2]{keV} peak from $^{60}$Co and above by the \unit[1238.1]{keV} peak from $^{214}$Bi.
The same ranges are used for the other decay modes.
Those choices allow that the spectrum in each ROI can be described by an extended p.d.f.\ $P(E|\mathbf{p})$\ consisting of a Gaussian signal peak and a linear background component: 
\begin{eqnarray}
P(E|\mathbf{p}) =
 \frac{\mu_S}{\sqrt{2\pi}\sigma}
 \cdot e^{\left(-\frac{(E-E_S)^2}{2\sigma^2}\right)}
 + b_0 + b_1\cdot E\ , \label{eqn:pdf}
\end{eqnarray}

where $E$\ is the energy and $\mathbf{p}$\ is the set of parameters consisting of
$E_S$\ being the signal peak position, $\mu_S$\ the signal count expectation,
$\sigma$\ the energy resolution and $b_0$ and $b_1$ the parameters describing the linear background polynomial.
The expectation of the signal counts is connected to the inverse half-life via
\begin{eqnarray}
\mu_S = \ln{2}\cdot \frac{N_A\cdot a_{74}\cdot M}{m_{\rm Se}}\cdot t\cdot \epsilon\cdot T_{1/2}^{-1}\ ,
\end{eqnarray}
where $N_A$\ is Avogadro's constant,
$a_{74}$\ the abundance of \sevs, $m_{\rm Se}$ the molar mass  of natural selenium,
$M$\ the mass of the selenium sample
and $\epsilon$\ the full \gray\ detection efficiency.
The probability of the data when combining both ROI is written as the product of Poisson probabilities of each bin:
\begin{eqnarray}
P(\mathbf{n}|\mathbf{p}) =
\prod \limits_{r=1}^{2} \prod \limits_{i=1}^{N_r} \frac{\lambda_{r,i}(\mathbf{p})^{n_{r,i}}}{n_{r,i}!} e^{-\lambda_{r,i}(\mathbf{p})}&&
\end{eqnarray}

where $\mathbf{n}$\ denotes the data, $r$\ the ROI,
$N_r$\ the number of bins in ROI $r$ and
$n_{r,i}$\ the number of counts in bin $i$\ of ROI $r$.
The expected number of counts $\lambda_{r,i}$\ in bin $i$\ of ROI $r$\ is obtained by integrating $P(E|\mathbf{p})$\ over the energy range of the bin.
By applying Bayes theorem the posterior probability $P(\mathbf{p}|\mathbf{n})$ is obtained which can be reduced to $P(T_{1/2}^{-1}|\mathbf{n})$ by integrating over the nuisance parameters.
Gaussian shaped prior probabilities are used for the efficiency, resolution and the peak positions to account for the systematic uncertainties of those parameters.
A flat prior is used for $b_0$, $b_1$ and $T_{1/2}^{-1}$ whereas the linear slope $b_1$ is restricted to negative values and the inverse half-life $T_{1/2}^{-1}$ to positive i.e. physical values.
In case of no signal, 90\% credibility lower half-life limits are obtained by the 90\% quantile of the marginalized posterior probability density of $T_{1/2}^{-1}$.



\subsection{Results}
\label{sec:results}
The results for all the considered decay modes are compiled in Tab.~\ref{tab:results}. Additionally, they are discussed in the following.

\paragraph{0$\nu$\ecec resonant to 2$^+_2$:}
Three \glines\ are expected from the two de-excitation branches of the $2^+_2$ state (Fig.~\ref{fig:decayscheme}). The \gline\ at \unit[608.4]{keV} is not considered for the analysis because of the $^{214}$Bi background \gline\ in its proximity.
The full energy detection efficiencies including the branching ratio are 1.9\% for the \unit[595.9]{keV} \gray\ and 0.7\% for the \unit[1204.2]{keV} \gray.

\begin{figure}
\centering
\includegraphics[width=0.6\textwidth]{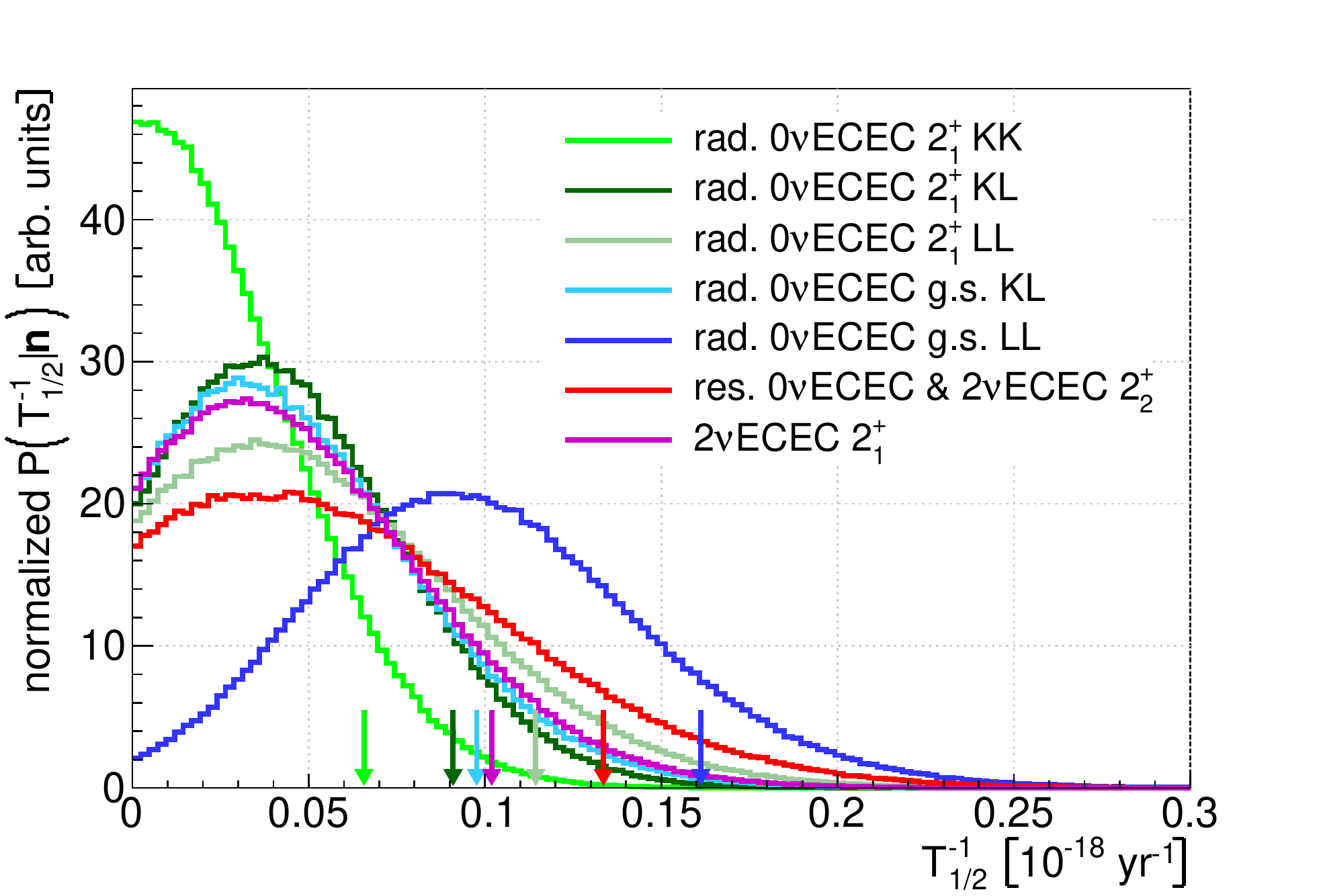}
\caption{%
Marginalized posterior probability density of the inverse half-life for all decay modes. The 90\% quantile of the distribution is indicated with vertical arrows and used to set the lower half-life limit.
\label{pic:posteriors}}
\end{figure}

\begin{figure}
\centering
\includegraphics[width=1.0\textwidth]{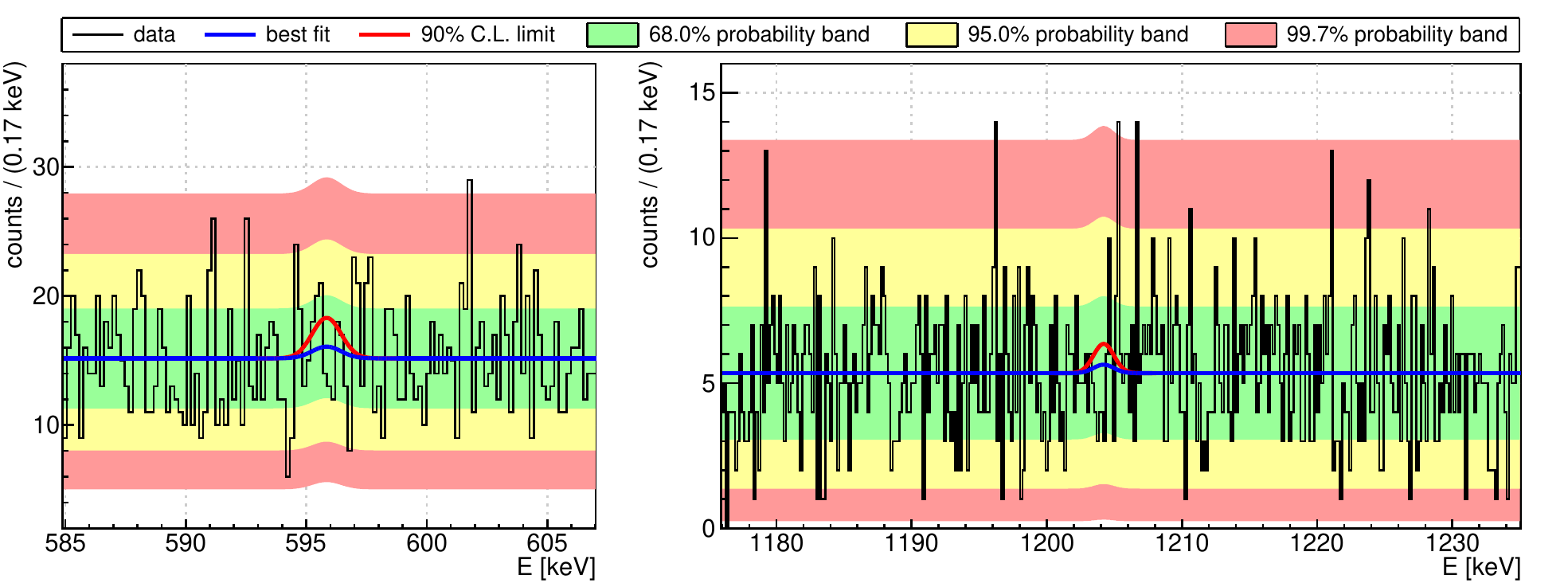}
\caption{%
Energy spectrum in the both regions of interest for the resonant transition into the $2^+_2$ state. Shown is the combined best fit (blue line) and the 90\% credibility limit (red line). The color bands show the central 68\%, 95\% and 99.7\% intervals for the Poisson probability of the counts in each bin with the expectation value given by the best fit.
\label{pic:fit}}
\end{figure}

The data shows a slight upward fluctuation of events at the position of the \unit[591.2]{keV} peak
and a downward fluctuation at the position of the \unit[1204.2]{keV} peak, when compared to the background level.
The maximum of the combined posterior probability $P(\mathbf{p}|\mathbf{n})$ is found at a value for $T_{1/2}^{-1}$
which corresponds to a signal count expectation of about 8 counts in the first ROI and about 3 counts in the second ROI
with background expectations of \unit[$88.3^{+2.3}_{-2.1}$]{cts/keV} and \unit[$31.2^{+0.8}_{-0.7}$]{cts/keV} respectively.
However, the zero signal case is included in the smallest 68\% interval of $P(T_{1/2}^{-1}|\mathbf{n})$ which is shown for this decay mode as well as for all other decay modes in Fig.~\ref{pic:posteriors}. The 90\% quantile used to calculate the lower half-life limit is indicated with vertical arrows. A lower half-life limit of \unit[$T_{1/2} > 0.70 \cdot 10^{19}$]{yr} (90\% credibility) is extracted for the 0$\nu$\ecec resonant decay of \sevs. The influence of the systematic uncertainties reduces the limit by about 1.3\%.
The energy spectrum around the ROI including the best fit and the limit are shown in Fig.~\ref{pic:fit}.



\begin{table}
\caption{\label{tab:results} Resulting half-life limits for the analyzed decay modes. Also given are the final state and the capture shells as well as the gamma energies expected from each decay. The energies in brackets are not included in the analysis, because they are inseparable from a background line.}
\begin{center}
\footnotesize
\begin{tabular}{l|cccc}
\hline
decay mode & final state & capture shells & \gray\ energies (efficiency) & limit (90\% cred.)\\
 & & & [keV] & [\unit[$10^{19}$]{yr}] \\
\hline
res. 0$\nu$\ecec    & 2$^+_2$  	& LL		& 595.85 (1.9\%), (608.35), 1204.21 (0.7\%)	& 0.70\\ 
rad. 0$\nu$\ecec	& g.s. 		& KL		& 1196.73 (2.2\%)					& 0.96\\ 
					& g.s. 		& LL		& 1206.42 (2.2\%)					& 0.58\\ 
					& 2$^+_1$ 	& KK		& 595.85 (2.7\%), 591.19 (2.7\%)	& 1.43\\ 
					& 2$^+_1$ 	& KL		& 595.85 (2.7\%), 600.88 (2.7\%)	& 1.03\\ 
					& 2$^+_1$ 	& LL		& 595.85 (2.7\%), (610.57)			& 0.82\\ 
2$\nu$\ecec			& 2$^+_1$ 	& KK,KL,LL	& 595.85 (3.0\%)					& 0.92\\ 
					& 2$^+_2$ 	& LL 		& 595.85 (1.9\%), (608.35), 1204.21 (0.9\%)	& 0.70\\ 
\hline
\end{tabular} 
\end{center}
\end{table}

\paragraph{0$\nu$\ecec radiative to ground state:}
One signal peak in the second ROI at \unit[1196.7]{keV} for the KL capture and \unit[1206.4]{keV} for the LL capture is expected from this decay mode.
The detection efficiency for this energy is 2.2\%.
The posterior probability $P(\mathbf{p}|\mathbf{n})$ peaks at a value corresponding to about 7 counts and 21 counts respectively.
For the LL capture, the zero signal case is outside of the smallest 95\% interval, but still inside the 99.7\% interval.
The probability for the zero signal case is still reasonably high enough so that it can not be rejected.
Lower half-life limits of \unit[$0.96\cdot 10^{19}$]{yr} (KL capture) and \unit[$0.58\cdot 10^{19}$]{yr} (LL capture) are obtained with 90\% credibility.
The latter limit is considerably lower because it is negatively affected by the upward fluctuation.
Both limits are about a factor 1.5 higher than previous limits of 
\unit[$0.64\cdot 10^{19}$]{yr} (KL) and \unit[$0.41\cdot 10^{19}$]{yr} (LL) \cite{bara07}.

\paragraph{0$\nu$\ecec radiative to $2^+_1$:}
This decay mode features a signal \gline\ at \unit[595.9]{keV} from the de-excitation \gray\ of the $2^+_1$ state.
Additionally, a second peak from the internal bremsstrahlung \gray\ at \unit[591.2]{keV} (KK), \unit[600.9]{keV} (KL capture) and \unit[610.6]{keV} (LL capture) is expected.
The \unit[610.6]{keV} \gline\ is inseparable from the $^{214}$Bi background \gline\ and is not used for the analysis.
The detection efficiency for these \grays\ is 2.7\%, which is higher than for the $2^+_2$ transitions, mainly because there is no branching from the $2^+_1$ state.
The maximum of $P(\mathbf{p}|\mathbf{n})$ corresponds to 0 counts for the signal expectation for the KK capture and about 11 counts per peak for both the KL and LL capture.
Lower half-life limits of \unit[$1.43\cdot 10^{19}$]{yr} (KK capture), \unit[$1.03\cdot 10^{19}$]{yr} (KL) and \unit[$0.82\cdot 10^{19}$]{yr} (LL)
are obtained. They are comparable to the previous limits of \unit[$1.57\cdot 10^{19}$]{yr} (KK), \unit[$1.12\cdot 10^{19}$]{yr} (KL) and \unit[$1.30\cdot 10^{19}$]{yr} (LL) reported in \cite{bara07}.

\paragraph{2$\nu$\ecec to $2^+_1$:}
Only one \unit[591.2]{keV} \gray\ from the de-excitation of the $2^+_1$ state is expected, with a relatively high detection efficiency of 3\% due to no branching and no summation effects.
The maximum of $P(\mathbf{p}|\mathbf{n})$ corresponds to an expectation of about 11 counts in the signal peak hinting at a slight upward fluctuation compared to the background.
A lower limit of \unit[$0.92\cdot 10^{19}$]{yr} could be extracted, which improves the previous best limit of \unit[$0.77\cdot 10^{19}$]{yr} \cite{bara07} by 16\%.

\paragraph{2$\nu$\ecec to $2^+_2$:}
The results for this decay mode are the same as for the resonant transition to the $2^+_2$ state, due to the identical signature of the de-excitation \gray\ cascade. This also applies for the best previous limit.\\

\section{Discussion}
The half-life limit for the 0$\nu$\ecec resonant transition found in this work is lower than the quoted limit of \unit[$1.5\cdot 10^{19}$]{yr} in Ref.~\cite{fre15}.
However, when comparing the experimental parameters, the current work has a substantially lower background level and a higher detection efficiency which indicates a significant improvement of sensitivity compared to Ref.~\cite{fre15}.
Even so the statistical extraction of the half-life limit  $T_{1/2}$ differs between the two analyses, 
a simple sensitivity estimates $T_{1/2}^{\rm sens.}$ of the measurements can be quantitatively compared based on only few experimental parameters.
Those parameters are the sample mass $m$, the measuring time $t$, the detection efficiency $\epsilon$ and the background rate $B$ (in units of counts per keV, day and kg sample mass) which are shown in Tab.~\ref{tab:comparison} for this work and the recent measurements in Ref.~\cite{bara07,fre11,fre15}
for the case of the \unit[595.9]{keV} \gline.
We re-evaluated the $T_{1/2}^{\rm sens.}$ of these measurements using the definition $T_{1/2}^{\rm sens.} =  R\cdot \alpha \cdot \epsilon \cdot \sqrt{m\cdot t}/\sqrt{B\cdot \Delta E}$ which is designed to compare different experiments \cite{AvignoneIII2005}.
$R$ is the \gline\ branching ratio and $\Delta E$ is the energy window of the peak window.
The isotope specific factor $\alpha$ is taken as \unit[$\ln{2}\cdot N_A \cdot a_{\rm Se74} \cdot /(1.6\cdot m_{\rm Se}) = 2.9\cdot10^{22}$]{kg$^{-1}$} with $N_A$ as Avogadro's constant, $a_{\rm Se74}$ the isotopic abundance of \sevs\ and $m_{\rm Se}$ the molar mass of natural selenium.
The factor of 1.6 is necessary for a confidence level of 90\%.
$T_{1/2}^{\rm sens.}$ and the quoted half-life limits are compared in the last two rows of Tab.~\ref{tab:comparison}.
In Ref.~\cite{bara07} a Bayesian approach is adopted including all three \glines\ for the 0$\nu$\ecec resonant transition.
The re-evaluated $T_{1/2}^{\rm sens.}$ is in agreement with the published limit.
In Ref.~\cite{fre11} the method of setting a limit is not given.
The published value is 3 orders of magnitude larger than $T_{1/2}^{\rm sens.}$ and thus unphysical.
The value was corrected in Ref.~\cite{fre15} but remains overestimated by a factor of 30 compared to $T_{1/2}^{\rm sens.}$.
In Ref.~\cite{fre15} a similar definition of the sensitivity as $T_{1/2}^{\rm sens.}$ is interpreted as the half-life limit.
We find again a factor of 20 discrepancy between the published value and the re-evaluated sensitivity.
The published limits cannot be explained by the statistical treatment of the data and are thus doubtful.
For this reason we reject the results from Ref.~\cite{fre11,fre15} and consider the obtained limit in the present work as experimentally more robust and trustworthy.
It is thus the currently most stringent limit for the 0$\nu$\ecec resonant transition of $^{74}$Se.


\begin{table}
\begin{threeparttable}
\caption{\label{tab:comparison}
Comparison of the parameters of the experiment for the \unit[595.9]{keV} \gline\ of the resonant transition into the $2^+_2$ state presented in this work with previous searches published in \cite{bara07,fre11,fre15}.
The background rates of \cite{bara07,fre11} were estimated from the spectra included in those publications. The sensitivities $T_{1/2}^{\rm sens.}$ are calculated as described in the text.
Limits and sensitivities are given at 90\% credibility/confidence level.}
\footnotesize\rm
\begin{tabular*}{\textwidth}{@{}l*{15}{@{\extracolsep{0pt plus12pt}}l}}
\br
 & quantity & symbol / unit & \cite{bara07} & \cite{fre11} &  \cite{fre15} & this work \\
\mr
& sample mass & $m$ [kg]
    & 0.563 & 3.0 & 3.0 & 2.5036 $^\mathrm{a}$ \\
& measuring time & $t$ [d]
    & 18.19 & 21.7 & 34 & 35.29 \\
& detection efficiency $^\mathrm{b}$ & $\epsilon$ [\%] 
    & 2.74 & 0.0091 \cite{fre15} & 0.884 & 2.70 \\
& background rate ($\pm 3\sigma$) & $B\cdot\Delta E\cdot m$ [d$^{-1}$]
    & $\sim$1 & $\sim$3 & 175 & $\sim$9 \\
\mr
& published limit & $T_{1/2}^{}$ [$10^{19}$~yr]
    & $>0.55$ $^\mathrm{c}$
    & $>4.3$ 
    & $>1.5$
    & $>0.75$ $^\mathrm{e}$\\
&&&& $>0.14$ $^\mathrm{d}$& & \\
& sensitivity & $T_{1/2}^{\rm sens.}$ [$10^{19}$~yr]
    & $0.35$
    & $0.0039$
    & $0.063$
    & $0.72$\\
\br
\end{tabular*}
\begin{tablenotes}
      \small
      \item $^\mathrm{a}$ a part of the same sample used in \cite{fre11,fre15}
      \item $^\mathrm{b}$ the branching ratio is not included in this efficiency
      \item $^\mathrm{c}$ Bayesian approach analyzing all three lines
      \item $^\mathrm{d}$ Erratum of \cite{fre11} value in Ref.\ \cite{fre15}
      \item $^\mathrm{e}$ Bayesian approach performing a spectral fit combining \unit[595.9]{keV} and \unit[1204.2]{keV} \glines\ (90\% credibility)
\end{tablenotes}
\end{threeparttable}
\end{table}

\section{Summary and conclusions}
A search for double electron captures of \sevs\ has been performed into the \unit[595.9]{keV} and \unit[1204.2]{keV} state as well as into the ground state of \gevs.
No significant signal was detected for any of the decay modes.
Lower half-life limits have been obtained which are up to a factor 1.5 larger than previous limits. The limit for the resonant decay to the \unit[1204.2]{keV} state was found as \unit[$0.70\cdot 10^{19}$]{yr} (90\% credibility). Apparently no resonance enhancement is visible. The realization of a resonance enhancement is anyhow strongly disfavored
by precision Q-value measurements using Penning-traps \cite{kol10,mou10}.
The obtained half-life limit is lower than the published results in Ref.~\cite{fre11,fre15};
however, severe issues have been identified with the analysis and the results are not used for comparison.
For this reason we consider the obtained limit in this paper more robust and trustworthy and the most stringent limit for $^{74}$Se.

\section{Acknowledgements}
The authors would like to thank D. Frekers for providing the Se sample.


\bigskip

\section*{References}

\begin{thebibliography}{10}

\bibitem{avi08} F. T. Avignone III et al., \Journal{\RMP}{80}{481}{2008}
\bibitem{Rodejohann:2011fr} W. Rodejohann, \Journal{\em Int. J. Mod. Phys. E}{20}{1833}{2011}
\bibitem{Rodejohann:2012cc} W. Rodejohann, \Journal{\em J. Phys. G}{39}{124008}{2012}

\bibitem{Schechter:1982} J. Schechter, J. W. F. Valle \Journal{\PRD}{25}{2951}{1982}
\bibitem{fon12} C.~S.~Fong, E.~Nardi, A.~Riotto, Adv. High Energy Phys. 2012, 158302 (2012)
\bibitem{gan13} A.~Gando et al.,  \Journal{\PRL}{110}{062502}{2013}
\bibitem{ago13} M.~Agostini et al., \Journal{\PRL}{111}{122503}{2013}
\bibitem{alb14} J.~B.~Albert et al., \Journal{\NAT}{510}{229}{2014}
\bibitem{hir94} M.~ Hirsch et al., \Journal{\ZPA}{347}{151}{1994}
\bibitem{Vergados:1983do} J. D. Vergados, \Journal{\NPB}{218}{109}{1983}
\bibitem{Doi:1993hj} M. Doi and T. Kotani, \Journal{Prog. Theo. Phys.}{89}{139}{1993}
\bibitem{Sujkowski:2004dub} Z. Sujkowski and S. Wycech, \Journal{\PRC}{70}{052501}{2004}
\bibitem{ber83} J. Bernabeu, A. de Rujula, C. Jarlskog   \Journal{\NPB}{223}{15}{1983}
\bibitem{Kotila:2014ira} J. Kotila, J. Barea, F. Iachello \Journal{\PRC}{89}{064319}{2014}


\bibitem{Nist:2014} A. Kramida, Yu. Ralchenko, J. Reader and {NIST ASD Team},
{\em NIST Atomic Spectra Database (version 5.2)} [Online], 2014

\bibitem{Krivoruchenko2011} M. I. Krivoruchenko, F. \^{S}imkovic, D. Frekers, A. Faessler,  \Journal{\NPA}{859}{140}{2011}



\bibitem{bara07} A. S. Barabash, Ph. Hubert, A. Nachab, V. Umatov,  \Journal{\NPA}{785}{371}{2007}
\bibitem{fre11} D. Frekers et al.,  \Journal{\NPA}{860}{1}{2011}
\bibitem{fre15} M. Jevskovsky et al., \Journal{\NIMA}{795}{268}{2015}


\bibitem{kol10} V. S. Kolhinen et al., \Journal{\PLB}{684}{17}{2010}
\bibitem{mou10} B. J. Mount, M. Redshaw, E. G. Myers, \Journal{\PRC}{81}{032501}{2010} 

\bibitem{NuclData06} B. Singh and A. R. Farhan, 
\Journal{\em Nuclear Data Sheets}{107}{1923}{2006}


\bibitem{Niese98} S. Niese, M. K\"ohler, B. Gleisberg, \Journal{{\em J. of Radioanalyt. and Nucl. Chem.}}{233}{167}{1998}
\bibitem{deg09} M. K\"{o}hler et al, \Journal{\IJARI}{67}{736}{2009}
\bibitem{degering08} D. Degering, M. K\"ohler, {\it IEEE Nucl. Sci. Sym. Dresden} {2008}
\bibitem{RGU} Report IAEA/RL/148, Vienna (1987)
\bibitem{Boswell:hc}M. Boswell, Y.-D. Chan, and J. Detwiler et al., \Journal{\em IEEE Trans. Nucl. Sci.}{58}{1212}{2011}
\bibitem{decay0} O.A. Ponkratenko, V. I. Tretyak , Yu. G. Zdesenko, \Journal{\PAN}{63}{1282}{2000}
\bibitem{BAT09} A. Caldwell, D. Koll\'ar, K. Kr\"oninger, \Journal{\em Comp. Phys. Commu.}{180}{2197}{2009}

\bibitem{AvignoneIII2005} F. T. Avignone III, G. S. King III and Yu G. Zdesenko, 
\Journal{\em New J. Phys.}{7}{6}{2005}




\end{thebibliography}
\end{document}